\documentclass[fleqn,twoside]{article}
\usepackage{espcrc2}

\title{Equilibrium current temperature quasi-oscillations in a
ferromagnetic loop}

\author{R.G. Mints\address{School of Physics and Astronomy, Raymond and
Beverly Sackler Faculty of Exact Sciences, Tel Aviv University,
Tel Aviv 69978, Israel}\thanks{Tel.: +~972-3-640-9165; fax:
+~972-3-642-2979. E-mail address: mints@ccsg.tau.ac.il.}}

\begin{document}

\begin{abstract}
Equilibrium persistent current carried by a small
ferromagnet-metal loop is considered. This current is shown to be
quasi periodic in temperature at low temperatures. The quasi
period is determined mainly by the temperature dependence of the
magnetization. \vspace{1pc}
\par
\noindent{\it PACS:} 73.23.Ra; 75.75.+a
\par
\noindent{\it Keywords:} Persistent current; Ferromagnet-metal
loop
\end{abstract}

\maketitle

\section{INTRODUCTION}
\par
An equilibrium persistent current arising in a static magnetic
field in a single normal-metal loop results in a magnetic
response, which periodically oscillates with the magnetic flux
$\phi$ threading the loop \cite{butt,imry}. These oscillations
have a fundamental period given by the flux quantum $\phi_0=hc/e$
and exist if the electron phase coherence is preserved. The
Josephson-type magnetic response of an isolated normal-metal loop
subjected to a static external magnetic field was predicted
theoretically \cite{butt} and studied experimentally for a variety
of mesoscopic systems: an array of about $10^7$ isolated
mesoscopic cooper rings \cite{levy}; a single, isolated
micron-size gold loop \cite{chan}; a GaAs-AlGaAs single mesoscopic
ring \cite{mail}; an array of about $10^5$ GaAs-AlGaAs single
mesoscopic rings \cite{reul}; and an array of 30 gold mesoscopic
rings \cite{jari}.
\par
In this paper we consider an equilibrium persistent current
carried by an isolated ferromagnet-metal ring in the absence of an
external magnetic field. We show that at low temperatures this
persistent current is quasi-periodic in temperature. The
quasi-period $\delta T$ is determined mainly by the temperature
dependence of the equilibrium magnetization of the ferromagnet.
\par
\section{FERROMAGNET-METAL RING}
\par
Consider a ferromagnet-metal ring at a certain temperature $T\ll
T_c$, where $T_c$ is the Courie temperature. Suppose that the
external magnetic field is equal to zero. In this case the
electrons of a ferromagnet metal are subjected to the internal
magnetic field ${\bf B}=4\pi{\bf M}$, where ${\bf M}$ is the
magnetization. In a small ferromagnet sample ${\bf M}$ is uniform
as the formation of magnetic domains increases the free energy
\cite{land}. Therefore, a magnetic flux $\phi$ is threading a
small ferromagnet-metal ring even in the absence of an external
magnetic field.
\par
A monotonic variation of the magnetic field $4\pi {\bf M}$ results
in a periodic in the flux $\phi$ equilibrium persistent current
oscillations in a ferromagnet-metal ring. This effect is similar
to the oscillations in a normal-metal ring subjected to an
external magnetic field. The flux $\phi$ induced by the field
${\bf B}=4\pi {\bf M}$ can be presented as $\phi=4\pi MA_{\rm
eff},$ where $A_{\rm eff}$ is an effective area of the ring. The
value of $A_{\rm eff}$ depends on the orientation of the
magnetization ${\bf M}$ and the specific geometry of the ring. In
particular, if {\bf M} is parallel to the axis of symmetry, then
$A_{\rm eff} \sim\pi dD$, where $d$ is the thickness and $D$ is
the diameter of the ring.
\par
The magnetization $M(T)$ of a ferromagnet is a nonlinear function
of the temperature $T$. A small variation of the temperature
$\Delta T\ll T$ results in a flux variation $\Delta\phi =4\pi
A_{\rm eff}\vert dM/dT\vert\Delta T$. The equilibrium persistent
current is periodic in $\phi$ with the period $\phi_0$. Therefore,
the nonlinear dependence $M(T)$ results in an equilibrium current
quasi-periodic in temperature. The quasi-period $\delta T$ follows
from the relation $\Delta\phi = \phi_0$, which leads to the
following expression
\begin{equation}
\delta T={\phi_0\over\strut{4\pi A_{\rm eff}\left\vert
{\displaystyle dM\over\strut\displaystyle dT} \right\vert}}.
\label{eq1}
\end{equation}
It is worth noting that Eq. (\ref{eq1}) is valid if $\delta T\ll
T$.
\par
At low temperatures the dependence $M(T)$ is given by the Bloch
law:
\begin{equation}
M=M_0\left[ 1-\alpha\left({T\over T_c}\right)^{3/2}\right],
\label{eq2}
\end{equation}
\noindent where $M_0$ is the saturation magnetization and the
constant $\alpha=0.2-0.5$ depending on the ferromagnet. It follows
from the Eq.~(\ref{eq2}) that \par
\begin{equation}
\left\vert{dM\over dT}\right\vert ={3\alpha M_0 \over 2T_c}
\left({T\over T_c}\right)^{1/2}.
\label{eq3}
\end{equation}
Combining the Eqs.~(\ref{eq1}) and (\ref{eq3}) we find for the
quasi-period $\delta T$ the final expression
\begin{equation}
\delta T={\phi_0 T_c\over 6\pi \alpha A_{\rm eff}M_0}
\left({T_c\over T}\right)^{1/2}.
\label{eq4}
\end{equation}
\par
\section{SUMMARY}
\par
To summarize, we demonstrate that the magnetic response of a
single one-domain ring of a ferromagnet-metal is quasi-periodic in
temperature even in the absence of an applied magnetic field. To
estimate the quasi-period $\delta T$ let us consider a small
dysprosium ring. Suppose the temperature $T=4.2\,$K and effective
area $A_{\rm eff}=3\times 10^{-8}\,$cm$^2$. Using for dysprosium
\cite{ferr} the data $T_c=89\,$K and $M_0=0.29\,$T and estimating
$\alpha\approx 0.3$ we find $\delta T=0.3\,$K. This value seems to
be reasonable for an experimental observation.
\par
\section{ACKNOWLEDGMENTS}
\par
I would like to thank M. Azbel, D. Khmelnitskii and A. Larkin for
useful discussions. This research is supported by grant
No.~2000-011 from the United States - Israel Binational Science
Foundation (BSF), Jerusalem, Israel.
\par

\end{document}